\documentclass[10pt]{iopart}
\usepackage{latexsym}
\usepackage{amsfonts}
\usepackage{amsbsy}
\usepackage{amssymb}   
\usepackage{accents}
\usepackage{hyperref}
\hypersetup{colorlinks=true,linkcolor=blue,urlcolor=blue,citecolor=blue}

\usepackage[makeroom]{cancel}

\usepackage{tensor}
\usepackage{color}
\usepackage{cite}

\newcommand{\intprod}{\mbox{$\; \put(0,0){\line(1,0){.9}}
		\put(.9,0){\line(0,1){1.6}} \; \, \,  $}}

\begin{document}

\title[The gauge symmetries of first-order general relativity with matter fields]
{The gauge symmetries of first-order general relativity with matter fields}

\author{Merced Montesinos$^{1}$, Diego Gonzalez$^{1,2}$ and Mariano Celada$^3$}

\address{$^1$ Departamento de F\'{\i}sica, Cinvestav, Avenida Instituto Polit\'ecnico Nacional 2508,
	San Pedro Zacatenco, 07360, Gustavo A. Madero, Ciudad de M\'exico, M\'exico.}
\address{$^2$ Instituto de Ciencias Nucleares, Universidad Nacional Aut\'onoma de M\'exico, Apartado Postal 70-543, Ciudad de M\'exico, 04510, M\'exico}
\address{$^3$ Departamento de F\'isica, Universidad Aut\'onoma Metropolitana Iztapalapa, San Rafael Atlixco 186, 09340 Ciudad de M\'exico, M\'exico.}

\eads{\mailto{merced@fis.cinvestav.mx},  \mailto{diego.gonzalez@correo.nucleares.unam.mx} and \mailto{mcelada@fis.cinvestav.mx}}


\begin{abstract}

In $n$-dimensional spacetimes  ($n>3$), there exists an internal gauge symmetry of the Palatini action with a cosmological constant that is the natural generalization of the so-called ``local translations'' of three-dimensional general relativity. We report the extension of this symmetry to include the minimal coupling of Yang-Mills and fermion fields to the Palatini action with a cosmological constant. We show that, as in the case of three-dimensional local translations, the extended symmetry depends on the energy-momentum tensor of the corresponding matter field and, for  fermions, it contains an additional term that in four dimensions is proportional to the axial fermion current. We also report the extension of the analog of this internal gauge symmetry for the Holst action with a cosmological constant by incorporating minimally coupled scalar and Yang-Mills fields, as well as a non-minimally coupled fermion field. In the last case, the extended symmetry is affected by both the Immirzi parameter and the energy-momentum tensor and, for fermions, it also depends on the axial fermion current.

\end{abstract}

\section{Introduction}

In the first-order formalism, the whole gauge symmetry of $n$-dimensional general relativity  ($n\geq 3$) can be captured by the set of symmetries composed of local Lorentz transformations and diffeomorphisms. However, even though diffeomorphism invariance is the symmetry at the heart of general relativity, it is not an exclusive property of general relativity (see for instance~\cite{Sotiriou2010}). In fact, diffeomorphisms are insensitive to the form of the action dictating the dynamics of the system and then, in this sense, the invariance under them constitutes a \textit{kinematical} symmetry~\cite{Utiyama,Thiebook}. On the other hand, in the three-dimensional setting the full symmetry of general relativity can alternatively be described by local Lorentz transformations and three-dimensional local translations~\cite{AchTow,Witten} (see also~\cite{Carlip2+1}). In this framework,  diffeomorphism invariance is no longer regarded as a fundamental symmetry, but as a derived symmetry that can be constructed out of these two symmetries. A particular feature of three-dimensional local translations is that they are sensitive to the presence of the cosmological constant, in contrast to diffeomorphisms. More remarkably, the idea of adopting three-dimensional local translations as a fundamental symmetry has proved to be useful for attacking the problem of quantizing  three-dimensional gravity~\cite{AchTow,Witten}. Motivated by these facts, in~\cite{MGCD2017} was uncovered the higher-dimensional generalization of three-dimensional local translations through the application of the converse of Noether's second theorem~\cite{Noether,Bessel-Hagen,EmmyNoether} to the first-order formulation of general relativity. One advantage of the obtained generalization is that it, together with local Lorentz transformations, offers a new way of reformulating the full symmetry of general relativity in terms of internal gauge symmetries. Moreover, such a generalization is affected by both the spacetime dimension and the structure of the Lagrangian underlying general relativity. This last property is evident in four-dimensional gravity where, instead of the Palatini action, we can use the Holst action~\cite{Holst}, in whose case the symmetry turns out to depend on the Immirzi parameter~\cite{MGCD2017}. A first step to gain insights into how the generalization of three-dimensional local translations is modified by the coupling of matter fields was given in~\cite{MGCD2017}, where it was shown that the non-minimal coupling of a scalar field to gravity in $n$ dimensions modifies the structure of this symmetry in a nontrivial way.

The purpose of this paper is twofold. First, we report the extension of the higher-dimensional generalization of three-dimensional local translations to include the minimal coupling of Yang-Mills and fermion fields to the $n$-dimensional Palatini action with a cosmological constant. In order to carry this out, we will follow the same procedure used in~\cite{MGCD2017}; that is, for each one of the considered matter-gravity couplings we will construct a Noether identity that, according to the converse of Noether's second theorem, allows us to determine an infinitesimal gauge symmetry of the theory. We will see that the obtained extended symmetry depends on the energy-momentum tensor of the corresponding matter field, and that for the particular case of fermions, it contains an additional term that in four dimensions is proportional to the axial fermion current. This despite the fact that the energy-momentum tensor for the Yang-Mills field is symmetric, whereas the one for the fermion field is nonsymmetric. The second purpose, which is related to the first one, is to report the extension of the analog of this gauge symmetry for the Holst action with a cosmological constant by encompassing minimally coupled scalar and Yang-Mills fields, and the one-parameter family of coupled fermions proposed by Mercuri in~\cite{Mercuri2006}. We find that, despite the presence of the Immirzi parameter, the modifications introduced by the matter couplings in the structure of the resulting extended symmetry are still modulated by the energy-momentum tensor, whereas the additional fermion field contributions are also proportional to the axial fermion current.


\section{Symmetries of the $3$-dimensional Palatini action} \label{Sec-3d}

It is instructive to first consider general relativity in three dimensions, where the basic strategy involved in the use of the converse of Noether's second theorem to uncover gauge symmetries can be illustrated more directly and neatly, without the technical difficulties arising in higher dimensions. 

The fundamental variables to describe the gravitational field in the first-order formalism are an orthonormal frame and a connection   compatible with the metric. Let $\mathcal{M}^3$ be an orientable three-dimensional manifold, and let $SO(\sigma)$ be the frame rotation group, with $SO(+1)=SO(3)$  for the Euclidean case ($\sigma=1$) and $SO(-1)=SO(1,2)$ for the Lorentzian one ($\sigma=-1$). General relativity with a cosmological constant is described by the Palatini action $S[e,\omega]= \int_{\mathcal{M}^3} L_{\rm Palatini}$, whose Lagrangian 3-form in terms of these variables is
\begin{eqnarray}\label{palatini3d}
L_{\rm Palatini} =\kappa  \epsilon_{IJK} e^I \wedge \left( R^{JK}[\omega] - \frac{\Lambda}{3} e^J \wedge e^K\right),
\end{eqnarray}
where $e^I$ is an orthonormal frame of 1-forms, $\omega_{IJ}=-\omega_{JI}$ is a Lorentz connection\footnote{Given that we are considering Euclidean and Lorentzian signatures at once, here and in the subsequent sections the word ``Lorentz''  refers to both signatures.} with curvature $R^I{}_J[\omega]:= d \omega^I{}_J + \omega^I{}_K \wedge \omega^K{}_J$, $\kappa$ is a constant related to Newton's constant, $\Lambda$ is the cosmological constant, and the Lorentz-invariant tensor $\epsilon_{IJK}$ is totally antisymmetric and such that $\epsilon_{012}=1$. The Latin indices $I, J,\dots$ denote Lorentz indices and, as such, are raised and lowered with the metric $(\eta_{IJ})=\mbox{diag} (\sigma,1,1)$. The variational derivatives of the action defined by (\ref{palatini3d}) are
\numparts
\begin{eqnarray}
{\mathcal E}_I :=\frac{\delta S}{\delta e^I}=\kappa\epsilon_{IJK} \left ( R^{JK} - \Lambda e^J \wedge e^K \right), \label{Pal3d-Eqmot1}\\
{\mathcal E}_{IJ} :=\frac{\delta S}{\delta \omega^{IJ}}=\kappa \epsilon_{IJK} D e^K, \label{Pal3d-Eqmot2}
\end{eqnarray}
\endnumparts
with $D$ being the Lorentz-covariant derivative defined by $\omega^I{}_J$. The three-dimensional Einstein's equations with a cosmological constant follow from (\ref{Pal3d-Eqmot1}) and (\ref{Pal3d-Eqmot2}) by setting ${\mathcal E}_I =0$ and ${\mathcal E}_{IJ}=0$. However, for the purposes of the present section, the variational derivatives ${\mathcal E}_I$ and ${\mathcal E}_{IJ}$ do not vanish in general.

The full gauge symmetry of the three-dimensional Palatini action can be described by (infinitesimal) local Lorentz transformations
\begin{eqnarray}
\delta_{\tau}e^I =\tau^I{}_{J} e^J, \nonumber\\
\delta_{\tau}\omega^{IJ}=- D \tau^{IJ}=-\left(d\tau^{IJ}+\omega^I{}_K \tau^{KJ} +\omega^J{}_K \tau^{IK} \right), \label{Pal3d-Lorentz}
\end{eqnarray}
and ``local translations''~\cite{Carlip2+1} (see, of course,~\cite{AchTow,Witten})
\begin{eqnarray}
\delta_{\rho} e^I = D \rho^I = d\rho^I+\omega^I{}_J\rho^J, \nonumber\\
\delta_{\rho} \omega^{IJ} =2\Lambda\rho^{[I} e^{J]}, \label{Pal3d-trans}
\end{eqnarray}
where the functions $\tau^{IJ}(=-\tau^{JI})$ and $\rho^I$ are the respective gauge parameters associated to these transformations. We recall that an infinitesimal transformation of the fields depending on arbitrary functions is a gauge symmetry of the action if the Lagrangian is quasi-invariant (invariant up to a total derivative) under it. For instance, the Lagrangian~(\ref{palatini3d}) is invariant under local Lorentz transformations (\ref{Pal3d-Lorentz})
\begin{eqnarray}
\delta_{\tau} L_{\rm Palatini} =0,
\end{eqnarray} 
and quasi-invariant under local translations (\ref{Pal3d-trans})
\begin{eqnarray}
\delta_{\rho} L_{\rm Palatini} = d\left[ \kappa \epsilon_{IJK} \rho^I \left( R^{JK}  + \Lambda e^J \wedge e^K \right)\right].  
\end{eqnarray}  		
On the other hand, the algebra of the transformations $\delta_{\tau}$ and $\delta_{\rho}$\footnote{Although the gauge parameters may in principle depend on the dynamical fields of the theory, throughout this paper we assume field-independent gauge parameters when computing the commutators of the gauge transformations.} closes
\begin{eqnarray}\label{algebra3d1}
\left [ \delta_{\tau_1}, \delta_{\tau_2} \right ] = \delta_{\tau_3} \qquad &&(\tau_3^{IJ}:=2\tau_1^{[I|K}  \tau_2^{|J]}{}_K ), \nonumber\\
\left [ \delta_{\rho}, \delta_{\tau} \right ] = \delta_{\rho_1} \qquad &&(\rho_1^I:=\tau^I{}_J\rho^J), \nonumber\\
\left [ \delta_{\rho_1}, \delta_{\rho_2} \right ] = \delta_{\tau} \qquad &&( \tau^{IJ} := 2 \Lambda \rho^{[I}_1 \rho^{J]}_2 ),
\end{eqnarray}
which can be recognized as the Lie algebra of the de Sitter group $SO(1,3)$ for $\Lambda > 0$, the anti-de Sitter group $SO(2,2)$ for $\Lambda<0$, or the Poincar\'e group $ISO(1,2)$ for $\Lambda=0$.

Apart from these symmetries, it is also well-known that diffeomorphisms are symmetries of the three-dimensional Palatini action. Infinitesimally, diffeomorphisms of $e^I$ and $\omega^{IJ}$ read, respectively,
\begin{eqnarray}
\delta_{\xi} e^I = {\mathcal L}_{\xi} e^I, \nonumber\\
\delta_{\xi} \omega^{IJ}= {\mathcal L}_{\xi} \omega^{IJ}, \label{Pal3d-diff}
\end{eqnarray}
where ${\mathcal L}_{\xi}$ is the Lie derivative along the vector field $\xi$. However, in this setting diffeomorphism invariance {\it is not} an independent symmetry. Indeed, by using Cartan's formula, namely, 
\begin{eqnarray}
	{\mathcal L}_{\xi} Q = d(\xi \intprod Q) + \xi \intprod dQ,
\end{eqnarray}
where $Q$ is an arbitrary $p$-form and ``$\intprod$'' stands for  contraction, the right-hand side (r.h.s.) of Eqs.~(\ref{Pal3d-diff}) can be written as
\begin{eqnarray}\label{difeos}
	{\mathcal L}_{\xi} e^I = (\delta_{\tau}+\delta_{\rho})e^I+\xi \intprod \left( \frac{\sigma}{2\kappa} \epsilon^{IJK}  {\mathcal E}_{JK}  \right), \nonumber\\
	{\mathcal L}_{\xi} \omega^{IJ} = (\delta_{\tau}+\delta_{\rho})\omega^{IJ}+ \xi \intprod \left( \frac{\sigma}{2\kappa} \epsilon^{IJK}  {\mathcal E}_{K}  \right), 
\end{eqnarray}
where $\tau^{IJ}= - \xi \, \intprod \omega^{IJ}$ and $\rho^I = \xi \, \intprod e^I$ are field-dependent gauge parameters. Thus, modulo terms involving the variational derivatives, infinitesimal diffeomorphisms can be constructed out of both local Lorentz transformations and local translations. 

Let us now review how, using the converse of Noether's second theorem, we can obtain local Lorentz transformations and three-dimensional local translations. The calculation is remarkably short. In order to get local Lorentz transformations we start by taking the covariant derivative of~(\ref{Pal3d-Eqmot2}), which leads to
\begin{eqnarray}
D {\mathcal E}_{IJ} = \kappa \epsilon_{IJK} D^2 e^K = \kappa \epsilon_{IJK} R^K{}_L \wedge e^L = e_{[I}  \wedge {\mathcal E}_{J]}, \nonumber
\end{eqnarray}
or 
\begin{eqnarray}\label{Pal3d-IdenL1}
D {\mathcal E}_{IJ} - e_{[I} \wedge {\mathcal E}_{J]}=0.
\end{eqnarray}
Next, we multiply (\ref{Pal3d-IdenL1}) by the gauge parameter $\tau^{IJ}(=-\tau^{JI})$ and rearrange it, thereby obtaining the off-shell identity
\begin{eqnarray}\label{Pal3d-IdenL2}
{\mathcal E}_I \wedge \underbrace{\tau^I{}_{J} e^J }_{\delta_{\tau} e^I}+ {\mathcal E}_{IJ} \wedge \underbrace{  (- D \tau^{IJ})}_{\delta_{\tau} \omega^{IJ}} + d \left( \tau^{IJ} {\mathcal E}_{IJ}\right)=0.
\end{eqnarray}	
By resorting to the converse of Noether's second theorem, the transformations~(\ref{Pal3d-Lorentz}) follow from the quantities multiplying the variational derivatives ${\mathcal E}_I$ and ${\mathcal E}_{IJ}$ in (\ref{Pal3d-IdenL2}). This in turn means that (\ref{Pal3d-IdenL1}) is the Noether identity associated to local Lorentz symmetry.

A completely analogous procedure can be used to obtain three-dimensional local translations. In this case, however, we start by computing the covariant derivative of~(\ref{Pal3d-Eqmot1}). Upon doing this and using the Bianchi identity $DR^{IJ}=0$, we get
\begin{eqnarray} 
D {\mathcal E}_I= 2\Lambda \kappa \epsilon_{IJK} e^J \wedge De^K = 2 \Lambda e^J \wedge \mathcal{E}_{IJ}, \nonumber
\end{eqnarray}
or
\begin{eqnarray} \label{Pal3d-Iden1}
	D {\mathcal E}_I -2 \Lambda e^J \wedge \mathcal{E}_{IJ}  =0.
\end{eqnarray}
Multiplying this equation by the gauge parameter $\rho^I$ and rearranging, we arrive at the off-shell identity
\begin{eqnarray}\label{Pal3d-Iden2}
{\mathcal E}_I \wedge \underbrace{D\rho^I }_{\delta_{\rho} e^I}+ {\mathcal E}_{IJ} \wedge \underbrace{  2\Lambda \rho^{[I} e^{J]}}_{\delta_{\rho} \omega^{IJ}} + d \left( - \rho^I {\mathcal E}_I\right)=0.
\end{eqnarray}
Appealing once again to the converse of Noether's second theorem, the transformations~(\ref{Pal3d-trans}) emerge from the quantities that multiply the variational derivatives in~(\ref{Pal3d-Iden2}). Moreover, it is now clear that~(\ref{Pal3d-Iden1}) is the Noether identity associated to three-dimensional local translations, which was reported in~\cite{MGCD2017}. 

The methodology outlined in the previous paragraphs was used in~\cite{MGCD2017} to reveal the higher-dimensional generalization of three-dimensional local translations and the analogous symmetry for the Holst action. In what follows, we are going to apply the same approach to extend the results of~\cite{MGCD2017} to include matter fields.

\section{Symmetries of the $n$-dimensional Palatini action with matter fields}

Before proceeding to consider the couplings of matter fields to gravity in $n$ dimensions, we would like to recall some facts about the higher-dimensional generalization of three-dimensional local translations, which will also allow us to establish some notation used in this section.

We consider an orientable $n$-dimensional manifold $\mathcal{M}^n$  (with $n\geq 3$) and also denote the frame rotation group by $SO(\sigma)$, with the convention that $SO(-1)=SO(1,n-1)$ and $SO(+1)=SO(n)$ for Lorentzian ($\sigma=-1$) and Euclidean ($\sigma=1$) manifolds, respectively; the internal indices $I,J,\dots$, which now run from $0$ to $n-1$, are raised and lowered with the internal metric $(\eta_{IJ})=\rm{diag}(\sigma,1,\dots,1)$. The Palatini action with a cosmological constant $\Lambda$ in $n$ dimensions  is given by $S[e,\omega]=\int_{\mathcal{M}^n} L_{\rm{Palatini}}$, with the Lagrangian $n$-form 
\begin{eqnarray}\label{palatinindim}
L_{\rm Palatini} = \kappa \left [ \star (e_I \wedge e_J) \wedge R^{IJ} [\omega] - 2 \Lambda \eta  \right ],
\end{eqnarray}
where $e^I$ is an orthonormal frame of 1-forms, $R^I{}_J[\omega]:= d \omega^I{}_J + \omega^I{}_K \wedge \omega^K{}_J=(1/2) \mathcal{R}^I{}_{JKL} e^K\wedge e^L$ is the curvature of  the  $SO(\sigma)$ connection 1-form  $\omega^I{}_J$, $\kappa$ is a constant (whose dimensions depend on $n$), $\eta:=(1/n!)\epsilon_{I_1\dots I_n} e^{I_1} \wedge \!\dots\!\wedge e^{I_n}$ is the volume form, and $\star$ is the Hodge dual operator
\begin{eqnarray} 
\star(e_{I_1} \wedge \dots\wedge e_{I_k}) =\frac{1}{(n-k)!} \epsilon_{I_1\dots I_kI_{k+1} \dots I_n} e^{I_{k+1}} \!\wedge \dots \wedge e^{I_n}. 
\end{eqnarray} 
Moreover, the internal tensor $\epsilon_{I_1\dots I_n}$ is totally antisymmetric and satisfies $\epsilon_{01\dots n-1}=1$. Notice that, for $n=3$, the Lagrangian (\ref{palatinindim}) collapses to (\ref{palatini3d}). The variational derivatives of the action defined by (\ref{palatinindim}) read
\numparts
\begin{eqnarray} 
\fl &&{\mathcal E}_I :=\frac{\delta S}{\delta e^I}= (-1)^{n-1} \kappa \star 
(e_I\wedge e_J\wedge e_K) 
\wedge \left [ R^{JK} [\omega] - \frac{2\Lambda}{(n-1)(n-2)} e^J \wedge e^K \right ], \label{moteqn1}\\
\fl &&{\mathcal E}_{IJ}:= \frac{\delta S}{\delta \omega^{IJ}}=(-1)^{n-1} \kappa \, D \star (e_I \wedge e_J),\label{moteqn2}
\end{eqnarray}
\endnumparts
where $D$ is the Lorentz covariant derivative. To obtain Einstein's equations with a cosmological constant we have to set both ${\mathcal E}_I$ and ${\mathcal E}_{IJ}$ equal to zero. However, throughout this paper, we assume that the variational derivatives are nonvanishing in general.

As is widely known, the full gauge symmetry of the $n$-dimensional Palatini action with a cosmological term is characterized by local Lorentz transformations and diffeomorphisms, which are still given by~(\ref{Pal3d-Lorentz}) and (\ref{Pal3d-diff}), respectively. It can be verified that the Lagrangian (\ref{palatinindim}) is invariant under infinitesimal local Lorentz transformations and is quasi-invariant under infinitesimal diffeomorphisms.

Recently, it was shown in~\cite{MGCD2017} that the infinitesimal internal gauge transformation
\begin{eqnarray}
\delta_{\rho} e^I = D\rho^I, \nonumber\\
\delta_{\rho} \omega^{IJ} = Z_n{}^{IJ}{}_{KL} \rho^{[K} e^{L]}, \label{gaugetrn}
\end{eqnarray}
where $\rho^I$ is the gauge parameter and
\begin{eqnarray}
Z_n{}^{IJ}{}_{KL}:=&&\frac{\sigma (n-3)}{(n-2)!} \Bigl( \epsilon^{IJNM_1 \dots M_{n-3}} \ast\!\mathcal{R}_{K M_1 \dots M_{n-3}NL} \nonumber\\
&&+ \ast\mathcal{R}\!\ast _{M_1 \dots M_{n-4}KL}{}^{M_1 \dots M_{n-4}IJ} \Bigr) + \frac{2 \Lambda}{n\!-\!2} \delta^{[I}_K \delta^{J]}_L, \label{Zn}
\end{eqnarray}
with
\numparts
\begin{eqnarray}
\ast\mathcal{R}_{I_1 \dots I_{n-2}MN} := \frac12  \epsilon_{I_1 \dots I_{n-2}KL} \mathcal{R}^{KL}{}_{MN},\\
\mathcal{R}\ast{}^{MN I_1 \dots I_{n-2}} := \frac12 \epsilon^{I_1 \dots I_{n-2}KL} \mathcal{R}^{MN}{}_{KL},
\end{eqnarray}
\endnumparts
is also a gauge symmetry of the $n$-dimensional Palatini action with a cosmological term. Certainly, the Lagrangian (\ref{palatinindim}) turns out to be quasi-invariant under this gauge transformation. It is worth mentioning that the gauge transformation (\ref{gaugetrn}) and local Lorentz transformations can be used for equivalently describing the whole gauge symmetry of general relativity~\cite{MGCD2017}. This follows from the fact that infinitesimal diffeomorphisms can be expressed as linear combinations of these two symmetries with field-dependent gauge parameters, modulo terms involving ${\mathcal E}_I$ and ${\mathcal E}_{IJ}$. 

Notice that, unlike local Lorentz transformations and diffeomorphisms, the structure of the internal gauge symmetry (\ref{gaugetrn}) depends explicitly on the spacetime dimension $n$; in particular, this property is exhibited by the transformation of the connection $\delta_{\rho} \omega^{IJ}$, which can be alternatively written as
\begin{eqnarray}\label{gaugetrn2}
\fl \delta_{\rho} \omega^{IJ} = \left[ C{}^{IJ}{}_{KL} - \frac{2(n-3)}{(n-2)} \delta^{[I}_K {\mathcal R}^{J]}{}_L \right] \rho^K e^L + \frac{1}{(n-2)} \left[  \frac{(n-3)}{(n-1)} {\mathcal R} + 2 \Lambda  \right] \rho^{[I} e^{J]},
\end{eqnarray}
where  ${\mathcal R}^{I}{}_J:={\mathcal R}^{KI}{}_{KJ}$ and ${\mathcal R}:={\mathcal R}^{I}{}_I$ are the Ricci tensor and the scalar curvature, respectively, whereas
\begin{eqnarray}
\fl C_{IJKL}&\equiv&R_{IJKL}-\frac{1}{(n-2)} \left(\eta_{IK} \mathcal{R}_{JL}-\eta_{JK} \mathcal{R}_{IL}+\eta_{JL} \mathcal{R}_{IK}-\eta_{IL} \mathcal{R}_{JK} \right) \nonumber\\
\fl &&+ \frac{1}{(n-1)(n-2)} \mathcal{R} \left(\eta_{IK}\eta_{JL}- \eta_{IL}\eta_{JK}  \right), \label{Weyl}
\end{eqnarray}
are the components of the Weyl tensor\footnote{Since we are working off-shell, $R_{IJKL}$ and $C_{IJKL}$ are not symmetric under the interchange of the first two indices with the last two, likewise the Ricci tensor is nonsymmetric in general.}~\cite{Gerardo}. To arrive at (\ref{gaugetrn2}) we just have to get rid of the Levi-Civita tensors in (\ref{Zn}) and then use (\ref{Weyl}). Remarkably, if we substitute $n=3$ into (\ref{gaugetrn2}), it turns out that $\delta_{\rho} \omega^{IJ}$ reduces \textit{off-shell} to that of three-dimensional local translations~(\ref{Pal3d-trans}). This means that the internal gauge symmetry~(\ref{gaugetrn}) is the natural higher-dimensional generalization of~(\ref{Pal3d-trans}). 

On the other hand, the commutators among $\delta_{\tau}$ and $\delta_{\rho}$ acting on both the frame and the connection lead to~\cite{MGCD2017}
\begin{eqnarray}
\left [ \delta_{\tau_1}, \delta_{\tau_2} \right ] = \delta_{\tau_3} \qquad (\tau_3^{IJ}:=2\tau_1^{[I|K}  \tau_2^{|J]}{}_K), \nonumber\\
\left [ \delta_{\rho}, \delta_{\tau} \right ] = \delta_{\rho_1} \qquad \hspace{2mm}(\rho_1^I:=\tau^I{}_J\rho^J), \nonumber\\
\left [ \delta_{\rho_1}, \delta_{\rho_2} \right ] = \delta_{\tau} + \mbox{terms involving ${\mathcal E}_I$ and ${\mathcal E}_{IJ}$}, \label{algebrapalatnd}
\end{eqnarray}
where $\tau^{IJ} := 2 Z_n{}_K{}^{[I}{}_L{}^{J]} \rho^{[K}_1 \rho^{L]}_2$ in the last line. As expected, for $n=3$ Eq.~(\ref{algebrapalatnd}) collapses to the algebra generated by local Lorentz transformations and three-dimensional local translations, namely (\ref{algebra3d1}). As a matter of check, we can see that in such a case $\tau^{IJ}=2 Z_3{}_K{}^{[I}{}_L{}^{J]} \rho^{[K}_1 \rho^{L]}_2= 2 \Lambda \rho^{[I}_1 \rho^{J]}_2$ and the terms involving ${\mathcal E}_I$ and ${\mathcal E}_{IJ}$ are absent. For $n>3$ the terms with ${\mathcal E}_I$ and ${\mathcal E}_{IJ}$ do not vanish, and hence the resulting commutator algebra (\ref{algebrapalatnd}) is open~\cite{Henneaux,Henneaux199047}. These facts  are also reflected at the level of the gauge identities satisfied by the commutators among $\delta_{\tau}$ and $\delta_{\rho}$, namely
\numparts
\begin{eqnarray}
\fl  {\mathcal E}_I \wedge \left [ \delta_{\tau_1}, \delta_{\tau_2} \right ] e^I + {\mathcal E}_{IJ} \wedge \left [ \delta_{\tau_1}, \delta_{\tau_2} \right ] \omega^{IJ}  +  d \left[ (-1)^{n-1} \tau_3^{IJ} {\mathcal E}_{IJ} \right] = 0 \quad (\tau_3^{IJ}=2\tau_1^{[I|K}  \tau_2^{|J]}{}_K), \nonumber\\ \label{GR-Id-LL}\\
\fl {\mathcal E}_I \wedge \left [ \delta_{\tau}, \delta_{\rho} \right ] e^I + {\mathcal E}_{IJ} \wedge \left [ \delta_{\tau}, \delta_{\rho} \right ] \omega^{IJ}  +  d\left[ (-1)^n  \rho_1^I {\mathcal E}_I  \right] = 0 \qquad (\rho_1^I=\tau^I{}_J\rho^J), \label{GR-Id-LN}\\
\fl {\mathcal E}_I \wedge \left [ \delta_{\rho_1}, \delta_{\rho_2} \right ] e^I + {\mathcal E}_{IJ} \wedge \left [ \delta_{\rho_1}, \delta_{\rho_2} \right ] \omega^{IJ}  +  d \left[ (-1)^{n-1} \tau^{IJ} {\mathcal E}_{IJ} + \Theta \right]= 0,  \label{GR-Id-NN}
\end{eqnarray}
where $\tau^{IJ}$ in (\ref{GR-Id-NN}) is the same as in the last line of (\ref{algebrapalatnd}), and 
\endnumparts
\begin{eqnarray}
\fl \Theta := \left. \frac{2 \kappa}{(n-3)!} \rho^{[M}_1 \rho^{N]}_2 \Big[ \left( \epsilon_{IJSKL_1 \dots L_{n-4}} Z_n{}_M{}^{I}{}_N{}^{J} + (n-3) \epsilon_{IJMKL_1 \dots L_{n-4}} Z_n{}^{IJ}{}_{NS} \right. \Big. \right. \nonumber \\
\left. \left. \left. + \Lambda \epsilon_{MNKSL_1 \dots L_{n-4}} \right) e^S \wedge e^{L_{1}} \wedge \dots \wedge e^{I_{n-4}}  \right. \right. \nonumber \\
\Big. - (n-3)(n-4)   \ast\! R_{MNK L_1 \dots L_{n-5} } \wedge e^{L_{1}} \wedge \dots \wedge e^{I_{n-5}}   \Big]  \wedge De^K. \nonumber
\end{eqnarray}
Since the commutator of two local Lorentz transformations is again a local Lorentz transformation, Eq.~(\ref{GR-Id-LL}) is identified with the gauge identity of a local Lorentz transformation. Similarly, because the commutator of a local Lorentz transformation and~(\ref{gaugetrn}) is a transformation of the same type as~(\ref{gaugetrn}), Eq.~(\ref{GR-Id-LN}) can be recognized as the gauge identity of a transformation of the same type as~(\ref{gaugetrn}). Finally, we turn to Eq.~(\ref{GR-Id-NN}) for which we consider the cases $n=3$ and $n>3$ separately. For $n=3$, we have that $\Theta=0$ and then~(\ref{GR-Id-NN}) reduces to
\begin{eqnarray}
 {\mathcal E}_I \wedge \left [ \delta_{\rho_1}, \delta_{\rho_2} \right ] e^I + {\mathcal E}_{IJ} \wedge \left [ \delta_{\rho_1}, \delta_{\rho_2} \right ] \omega^{IJ}  +  d\left( \tau^{IJ} {\mathcal E}_{IJ} \right)=0,\label{GR3-Id}
\end{eqnarray}
which is the gauge identity of a local Lorentz transformation with gauge parameter $\tau^{IJ}=2 \Lambda \rho^{[I}_1 \rho^{J]}_2$, reflecting the fact that the commutator of two transformations (\ref{gaugetrn}) (or (\ref{Pal3d-trans})) is a local Lorentz transformation. For $n>3$, the terms inside the exterior derivative of Eq.~(\ref{GR-Id-NN}) are a contribution of a local Lorentz transformation plus terms proportional to the variational derivatives.

Another interesting feature of the internal gauge symmetry~(\ref{gaugetrn}) is that it is also sensitive to the presence of matter fields. This was first explored in~\cite{MGCD2017}, where it was considered the non-minimal coupling of a scalar field to gravity in $n$ dimensions. In this section, we will focus on extending the internal gauge transformation~(\ref{gaugetrn}) by including the minimal couplings of Yang-Mills and fermion fields to general relativity.

\subsection{Yang-Mills field}\label{sec-nPal-YM}

The minimal coupling of a Yang-Mills field to the $n$-dimensional Palatini Lagrangian with a cosmological constant is accomplished by
 \begin{eqnarray}\label{nPal-YM-Lag}
 L = L_{\rm Palatini} + \alpha \star F_a \wedge F^a,
 \end{eqnarray}
where $F^a:=dA^a+(1/2) f^a{}_{bc}A^b\wedge A^c=(1/2)F^a{}_{KL} e^K\wedge e^L$ is the field strength of the Yang-Mills connection~$A^a$ and $\alpha$ is a real parameter. The internal indices $a,b,\dots$  are raised and lowered with the Killing-Cartan metric of the Lie algebra of the semisimple Yang-Mills gauge group, and we assume that the structure constants $f_{abc}$ are totally antisymmetric. The action defined by (\ref{nPal-YM-Lag}) possesses the following variational derivatives:
 \numparts
  \begin{eqnarray}
 \fl E_I := \frac{\delta S}{\delta e^I} =  {\mathcal E}_I  + (-1)^{n-1} T_{JI}\star e^J, \label{nPal-YM-Eqmot1} \\
  \fl E_{IJ} :=  \frac{\delta S}{\delta \omega^{IJ}} = {\mathcal E}_{IJ}, \label{nPal-YM-Eqmot2}\\
 \fl E_{a} := \frac{\delta S}{\delta A^a} = (-1)^{n-1} 2 \alpha \mathcal{D}\star F_a = (-1)^{n-1} 2 \alpha \left(d\star F_a+f_{abc}A^b \wedge \star F^c \right) , \label{nPal-YM-Eqmot3}
 \end{eqnarray}
 \endnumparts
 where ${\mathcal E}_I$ and ${\mathcal E}_{IJ}$ are given by (\ref{moteqn1}) and (\ref{moteqn2}), respectively, and 
 \begin{eqnarray}\label{nPal-YM-EMtensor}
 T_{IJ}:=-2\alpha\left( F_{aIK} F^{a}{}_{J}{}^{K} -\frac{1}{4} F_{aKL} F^{aKL}  \eta_{IJ}\right),
 \end{eqnarray}
is the conventional symmetric energy-momentum tensor for the Yang-Mills field. Here, $\mathcal{D}$ is the covariant derivative corresponding to the connection $A^a$. Notice that the space-time connection $\omega^I{}_J$ is still on-shell torsion-free.
 
The action defined by (\ref{nPal-YM-Lag}) is invariant under local Lorentz transformations, diffeomorphisms, and pure Yang-Mills transformations 
\begin{eqnarray}\label{nPal-YM-YM}
\delta_{\lambda} e^I=0, \qquad \delta_{\lambda}\omega^{IJ}=0, \qquad  \delta_{\lambda}A^a=\mathcal{D}\lambda^a,
\end{eqnarray} 
with gauge parameter $\lambda^a$. Furthermore, we will see that it also has an internal  gauge symmetry analogous to (\ref{gaugetrn}). In order to uncover such a symmetry, we follow the procedure described in section~\ref{Sec-3d} and take into account that (\ref{gaugetrn}) is the generalization of (\ref{Pal3d-trans}), which means that our first step is to compute $DE_I$. The covariant derivative of the first term on the r.h.s. of~(\ref{nPal-YM-Eqmot1}), after using the Bianchi identity, $DR^{IJ}=0$, gives
\begin{eqnarray}\label{nPal-YM-DEI1}
D\mathcal{E}_I=(-1)^{n-1} E_{KL} \wedge Z_n{}^{KL}{}_{IJ} e^J, 
\end{eqnarray}
where $Z_n{}^{IJ}{}_{KL}$ is given by (\ref{Zn}), while the covariant derivative of the second term leads to
\begin{eqnarray}\label{nPal-YM-DEI2}
D\left(T_{IJ}\star e^J\right) = E_{KL} \wedge Q_n{}^{KL}{}_{IJ} e^J+E_a \wedge F^a{}_{IJ}e^J,
\end{eqnarray}
with
\begin{eqnarray}\label{nPal-YM-Q}
Q_n{}^{IJ}{}_{KL} &:=& \alpha \kappa^{-1} \left( 2 F^{a[I|M} F_{aLM} \delta^{|J]}_K - \frac{2}{n-2} F^{a[I|M} F_{aKM} \delta^{|J]}_L  \right. \nonumber\\
&& \left.+ \frac{3}{2(n-2)} F^{aMN} F_{aMN} \delta^{[I}_K \delta^{J]}_L \right),
\end{eqnarray}
where we have used the Bianchi identity $\mathcal{D}F^a=0$. Using (\ref{nPal-YM-DEI1}) and (\ref{nPal-YM-DEI2}), our result for $DE_I$ is
\begin{eqnarray}
\fl DE_I = (-1)^{n-1} E_{KL} \wedge \left(Z_n{}^{KL}{}_{IJ} + Q_n{}^{KL}{}_{IJ} \right)e^J+ (-1)^{n-1} E_{a} \wedge F^a{}_{IJ} e^J, \nonumber
\end{eqnarray}
or 
\begin{eqnarray}\label{nPal-YM-IdSym}
\fl (-1)^n DE_I+E_{KL} \wedge \left(Z_n{}^{KL}{}_{IJ} + Q_n{}^{KL}{}_{IJ} \right)e^J+E_{a} \wedge F^a{}_{IJ} e^J=0.
\end{eqnarray}
Next, multiplying (\ref{nPal-YM-IdSym}) by the gauge parameter $\rho^I$ and rearranging, we end up with the off-shell identity:
\begin{eqnarray}\label{nPal-YM-IdSym2}
\fl E_I \! \wedge\! \underbrace{D\rho^I }_{\delta_{\rho} e^I}+ E_{IJ} \!\wedge\! \underbrace{\left(Z_n{}^{IJ}{}_{KL} + Q_n{}^{IJ}{}_{KL}\right)\rho^K e^L}_{\delta_{\rho} \omega^{IJ}} + E_{a} \!\wedge\! \underbrace{F^a{}_{IJ} \rho^I e^J }_{\delta_{\rho}A^a} + d \left [ (-1)^n \rho^I {\mathcal E}_I\right ]\!=\!0.
\end{eqnarray} 

According to the converse of Noether's second theorem, the quantities next to $E_I$, $E_{IJ}$, and $E_{a}$ in (\ref{nPal-YM-IdSym2}) correspond to the gauge transformation generated by the Noether identity~(\ref{nPal-YM-IdSym}). Therefore, from (\ref{nPal-YM-IdSym2}) we obtain the infinitesimal internal transformation
\begin{eqnarray}\label{nPal-YM-newtr}
& \delta_{\rho} e^I=D\rho^I,\nonumber\\
& \delta_{\rho}\omega^{IJ}=\left(Z_n{}^{IJ}{}_{KL}+Q_n{}^{IJ}{}_{KL}\right)\rho^K e^L,\nonumber\\
& \delta_{\rho}A^a=\rho \intprod F^a=F^a{}_{IJ}\rho^I e^J \qquad (\rho=\rho^I \partial_I),
\end{eqnarray} 
which is a gauge symmetry of the $n$-dimensional Palatini action with a cosmological constant and a minimally coupled Yang-Mills field. Here, $\partial_I$ is the dual basis of $e^I$ ($\partial_I\intprod e^J=\delta_I^J$). Actually, under the gauge transformation (\ref{nPal-YM-newtr}) the Lagrangian (\ref{nPal-YM-Lag}) is quasi-invariant because
\begin{eqnarray}\label{nPal-YM-changeL}
\fl \delta_{\rho} L &=& d\left\{\frac{\rho^I}{n-2}  \left[  \kappa\star 
(e_I\wedge e_J\wedge e_K)\wedge\left( R^{JK}  + \frac{2 \Lambda}{(n-1)(n-2)} e^J \wedge e^K \right) \right. \right. \nonumber \\  
\fl &&  \left. \left. + 2\alpha \left( -F_{aIK} F^{a}{}_{J}{}^K + \frac{3}{4} F_{aKL} F^{aKL} \eta_{IJ} \right) \star e^J \right]\right\}.
\end{eqnarray}

It is worth noting that the transformation of the frame in~(\ref{nPal-YM-newtr}) remains the same as in (\ref{gaugetrn}). However, this is no longer the case for the transformation of the spacetime connection, which gets modified by the presence of the Yang-Mills field. More precisely, the transformation of the connection in~(\ref{nPal-YM-newtr}), as compared with its counterpart in~(\ref{gaugetrn}), contains the new term $Q_n{}^{IJ}{}_{KL}$, which depends on the Yang-Mills field and can be recast in terms of the symmetric energy-momentum tensor (\ref{nPal-YM-EMtensor}) as
\begin{eqnarray}\label{nPal-YM-Q2}
Q_n{}^{IJ}{}_{KL} = \kappa^{-1} \left(  \delta^{[I}_K T^{J]}{}_L - \frac{1}{n-2} \delta^{[I}_L T^{J]}{}_K - \frac{1}{n-2} T^M{}_M \delta^{[I}_K \delta^{J]}_L \right).
\end{eqnarray}
Note that for $n=3$ the extra term (\ref{nPal-YM-Q2}) does not vanish, and hence even three-dimensional local translations are also affected by the presence of the Yang-Mills field. 


\subsection{Fermions}\label{sec-nPal-D} 

Another interesting kind of matter field worth considering is the fermionic one. Let us assume throughout this subsection a Lorentzian ($\sigma=-1$) manifold $\mathcal{M}^n$. By including a cosmological term, the action for the minimal coupling of fermions to gravity is $S[e,\omega,\psi]=\int_{\mathcal{M}^n}L$, with the Lagrangian $n$-form
\begin{eqnarray}\label{nPal-D-Lag}
 L = L_{\rm Palatini} + \frac{1}{2} \left(\overline{\psi}  \gamma^I D\psi - \overline{D\psi} \gamma^I \psi \right)\wedge \star e_I -m\overline{\psi} \psi \eta, 
\end{eqnarray}
where $\psi$ is the fermion field, $\overline{\psi}=\rmi \psi^\dagger \gamma^0$ denotes the Dirac adjoint with $\psi^\dagger$ the Hermitian adjoint of $\psi$, and $m$ is the fermion mass. Also, $\gamma^I$ are Dirac's gamma matrices satisfying the Clifford algebra relations $\{\gamma_I,\gamma_J\}=2\eta_{IJ} \mathbb{I}$ with $(\eta_{IJ})=\rm{diag}(-1,1,\dots,1)$ and $\mathbb{I}$ being the $2^{[n/2]}\times 2^{[n/2]}$ identity matrix\footnote{Recall that the spinor space in $n$ spacetime dimensions has complex dimension $2^{[n/2]}$, where [x] is the integer part of x.}, while the Lorentz-covariant derivative on $\psi$ and $\overline{\psi}$ is defined by
\numparts
\begin{eqnarray} 
D\psi := d\psi + \frac{1}{2} \omega^{IJ} \sigma_{IJ} \psi, \label{nPal-D-Covar1}\\
\overline{D\psi} := d\overline{\psi} - \frac{1}{2}\overline{\psi} \omega^{IJ} \sigma_{IJ}, \label{nPal-D-Covar2}
\end{eqnarray}
\endnumparts
where $\sigma_{IJ}:=(1/4) [\gamma_I,\gamma_J]$ are the generators of the spin representation of the Lie algebra of the Lorentz group. The variational derivatives coming from the action defined by (\ref{nPal-D-Lag}) are
\numparts
\begin{eqnarray}
\fl E_I := \frac{\delta S}{\delta e^I} = {\mathcal E}_I + (-1)^{n-1} T_{JI}\star e^J, \label{nPal-D-Eqmot1} \\
\fl E _{IJ} :=  \frac{\delta S}{\delta \omega^{IJ}} =  {\mathcal E}_{IJ} + \frac{(-1)^{n-1}}{4} \overline{\psi} \{ \gamma_K,\sigma_{IJ}\} \psi \star e^K  , \label{nPal-D-Eqmot2}\\
\fl E_{\psi} := \frac{\delta S}{\delta \psi} = -  \overline{D\psi} \, \gamma^I \wedge \star e_I - \frac{1}{2} \overline{\psi} \, \gamma^I D\star e_I - m \overline{\psi} \eta , \label{nPal-D-Eqmot3} \\
\fl E_{\overline{\psi}} := \frac{\delta S}{\delta \overline{\psi}} =  \gamma^I D\psi   \wedge \star e_I + \frac{1}{2} \gamma^I \psi \, D\star e_I - m \psi \eta , \label{nPal-D-Eqmot4}
\end{eqnarray}
\endnumparts
where ${\mathcal E}_I$ and ${\mathcal E}_{IJ}$ are given by (\ref{moteqn1}) and (\ref{moteqn2}), respectively, whereas
\begin{eqnarray}\label{nPal-D-EMtensor}
\fl T_{IJ}:=-\frac{1}{2} \left( \overline{\psi} \gamma_I D_J \psi - \overline{D_J \psi} \gamma_I \psi \right) + \frac12 \left(\overline{\psi} \gamma^K D_K \psi - \overline{D_K \psi} \gamma^K \psi  - 2m \overline{\psi} \psi\right) \eta_{IJ},
\end{eqnarray}
for $D_I\psi:=\partial_I\intprod D\psi$ and $\overline{D_I\psi}:=\partial_I\intprod \overline{D\psi}$, is the nonsymmetric energy-momentum tensor\footnote{Note that $T_{IJ}$ is nonsymmetric since $T_{(IJ)}=T_{IJ}+\frac{1}{2} \left( \overline{\psi} \gamma_{[I} D_{J]} \psi + \overline{D_{[I} \psi} \gamma_{J]} \psi \right)\neq T_{IJ}$.}. Notice that the equation of motion $E _{IJ}=0$ implies that the spacetime connection $\omega^I{}_J$ is no longer on-shell torsion-free. This is a consequence of the fact that $\omega^I{}_J$ is involved in the matter terms added to the Palatini action in~(\ref{nPal-D-Lag}), thus introducing torsion into the theory.

The aim here is to obtain the internal gauge symmetry of the same kind of (\ref{gaugetrn}) under which the Lagrangian (\ref{nPal-D-Lag}) is quasi-invariant.  Before continuing, let us recall that the action defined by (\ref{nPal-D-Lag}) is  invariant under both local Lorentz transformations and diffeomorphisms. 

By following a strategy along the same lines of the previous subsection, we begin by taking the covariant derivative of $E_I$. For the first term of the r.h.s. of (\ref{nPal-D-Eqmot1}) we have
\begin{eqnarray}\label{nPal-D-DEI1}
\fl D\mathcal{E}_I= (-1)^{n-1} E_{KL}\wedge Z_n{}^{KL}{}_{IJ}e^J+\frac{(-1)^n}{4} \overline{\psi} \{ \gamma^J,\sigma_{KL}\} \psi Z_n{}^{KL}{}_{IJ} \, \eta,
\end{eqnarray}
where we have used the Bianchi identity $DR^{IJ}=0$, whereas for the second term we get
\begin{eqnarray}\label{nPal-D-DEI2}
\fl D\left( T_{JI}\star e^J \right)&=& - E_J \wedge \left(\frac{1}{4\kappa} \overline{\psi} \{\gamma^J, \sigma_{IK}\} \psi e^K\right) + E_{KL}\wedge F_n{}^{KL}{}_{IJ}e^J \nonumber\\
\fl&&+E_{\psi} D_I \psi + \overline{D_I \psi} E_{\overline{\psi}}  + \frac{1}{4} \overline{\psi} \{ \gamma^J,\sigma_{KL}\} \psi Z_n{}^{KL}{}_{IJ} \, \eta,
\end{eqnarray}
with
\begin{eqnarray}\label{nPal-D-F}
\fl F_n{}^{IJ}{}_{KL} &=& \kappa^{-1} \left[ -\frac{1}{2} \delta^{[I}_{K} \left(  \overline{\psi} \gamma^{J]} D_L \psi - \overline{D_L \psi} \gamma^{J]} \psi \right) \right. \nonumber \\
\fl && \left. + \frac{1}{2(n-2)} \delta^{[I}_{L} \left(  \overline{\psi} \gamma^{J]} D_K \psi - \overline{D_K \psi} \gamma^{J]} \psi \right) + \frac{1}{(n-2)}   m \overline{\psi} \psi \delta^{[I}_K \delta^{J]}_L \right],
\end{eqnarray}
where we have used the relations $\{\sigma_{IJ},\sigma_{KL}\}=-2 (\eta_{[I|K}\sigma_{|J]L}-\eta_{[I|L}\sigma_{|J]K})$ and $\{\gamma^J,\sigma_{IJ}\}=0$. Using (\ref{nPal-D-DEI1}) and (\ref{nPal-D-DEI2}), we can easily write $DE_I$ as
\begin{eqnarray}
\fl DE_I &=& E_J \wedge \left[\frac{(-1)^{n} }{4\kappa} \overline{\psi} \{\gamma^J, \sigma_{IK}\} \psi e^K\right] + (-1)^{n-1}  E_{KL} \wedge \left(Z_n{}^{KL}{}_{IJ} - F_n{}^{KL}{}_{IJ} \right)e^J \nonumber \\
\fl  && + (-1)^{n-1}  E_{\psi} D_I \psi + (-1)^{n-1}  \overline{D_I \psi} E_{\overline{\psi}}, \nonumber
\end{eqnarray}
or
\begin{eqnarray}\label{nPal-D-IdSym}
 (-1)^{n} DE_I - E_J \wedge \left(\frac{1}{4\kappa} \overline{\psi} \{\gamma^J, \sigma_{IK}\} \psi e^K\right)  \nonumber \\
+ E_{KL} \wedge \left(Z_n{}^{KL}{}_{IJ} + F_n{}^{KL}{}_{IJ} \right)e^J + E_{\psi} D_I \psi + \overline{D_I \psi} E_{\overline{\psi}}   =0.
\end{eqnarray}
To proceed further, we multiply this equation by the parameter $\rho^I$ and rearrange it, which leads to 
\begin{eqnarray}\label{nPal-D-IdSym2}
\fl E_I \wedge \underbrace{\left(D\rho^I-\frac{1}{4\kappa} \overline{\psi} \{\gamma^I, \sigma_{JK}\} \psi \rho^J e^K\right) }_{\delta_{\rho} e^I}+ E_{IJ} \wedge \underbrace{\left(Z_n{}^{IJ}{}_{KL} + F_n{}^{IJ}{}_{KL}\right)\rho^K e^L}_{\delta_{\rho} \omega^{IJ}} \nonumber \\
\fl + E_{\psi} \underbrace{\rho^I D_I \psi}_{\delta_{\rho} \psi}  + \underbrace{ \rho^I \overline{D_I \psi} }_{\delta_{\rho} \overline{\psi}} E_{\overline{\psi}} + d \left [ (-1)^n \rho^I E_I\right ]=0.
\end{eqnarray}

Having the off-shell identity (\ref{nPal-D-IdSym2}), and bearing in mind the converse of Noether's second theorem, we can immediately read off the gauge transformation generated by the Noether identity~(\ref{nPal-D-IdSym}), namely
\begin{eqnarray}\label{nPal-D-newtr}
& \delta_{\rho} e^I=D\rho^I-\frac{1}{4\kappa} \overline{\psi} \{\gamma^I, \sigma_{JK}\} \psi \rho^J e^K,\nonumber\\
& \delta_{\rho}\omega^{IJ}=\left(Z_n{}^{IJ}{}_{KL}+F_n{}^{IJ}{}_{KL}\right)\rho^K e^L,\nonumber\\
& \delta_{\rho} \psi=\rho^I D_I \psi=\rho \intprod D\psi \qquad (\rho=\rho^I \partial_I), \nonumber\\
& \delta_{\rho} \overline{\psi}=\rho^I \overline{D_I \psi}=\rho \intprod \overline{D\psi}. 
\end{eqnarray} 
Equation~(\ref{nPal-D-newtr}) is a gauge symmetry of the $n$-dimensional Palatini action with a cosmological constant and minimally coupled fermion fields. As a matter of fact, the Lagrangian (\ref{nPal-D-Lag}) is quasi-invariant under~(\ref{nPal-D-newtr}) since
\begin{eqnarray}\label{nPal-D-changeL}
\fl \delta_{\rho} L &=& d\left\{\frac{\rho^I}{n-2}  \left[  \kappa\star 
(e_I\wedge e_J\wedge e_K)\wedge\left( R^{JK} + \frac{2 \Lambda}{(n-1)(n-2)} e^J \wedge e^K \right) \right. \right. \nonumber \\  
\fl &&  \left. \left. +T_{JI} \star e^J - \frac{1}{n-1} \left(T^J{}_J- (n-2) m\overline{\psi} \psi  \right) \star e_I \right]\right\}.
\end{eqnarray}

Some comments regarding the gauge transformation (\ref{nPal-D-newtr}) are appropriate. As far as the transformation of the frame is concerned, it is interesting to note that it now has an additional term as compared to that of (\ref{gaugetrn}). Such a term is a consequence of the fermionic matter and in the particular case of $n=4$ is proportional to the axial fermion current $J^I=\rmi \overline{\psi} \gamma^5 \gamma^I \psi$ with $\gamma^5:=\rmi \gamma^0 \gamma^1 \gamma^2 \gamma^3$ since $\{\gamma_I, \sigma_{JK}\}=\rmi \epsilon_{IJKL} \gamma^5 \gamma^L$. With regard to the transformation of the connection, it is not surprising, at this point, that its structure gets modified by the presence of fermions, which show up in the extra term $F_n{}^{IJ}{}_{KL}$. In addition, it is not difficult to realize that $F_n{}^{IJ}{}_{KL}$ can also be expressed in terms of the nonsymmetric energy-momentum tensor~(\ref{nPal-D-EMtensor}) as
\begin{eqnarray}\label{nPal-D-F2}
F_n{}^{IJ}{}_{KL} = \kappa^{-1}\! \! \left( \delta^{[I}{}_K T^{J]}{}_L - \frac{1}{n-2} \delta^{[I}_L T^{J]}{}_K - \frac{1}{n-2} T^M{}_M \delta^{[I}_K \delta^{J]}_L \right). 
\end{eqnarray}
Thus, the transformations of the connection in (\ref{nPal-YM-newtr}) and (\ref{nPal-D-newtr}) have exactly the same form when written in terms of the respective energy-momentum tensors. We stress, however, that while the energy-momentum tensor in~(\ref{nPal-YM-Q2}) is symmetric, the energy-momentum tensor in~(\ref{nPal-D-F2}) is not.

\section{Symmetries of the Holst action with matter fields}

General relativity in four dimensions can be also described by the Holst action~\cite{Holst}, which has served as the starting point of  the loop approach to quantum gravity~\cite{RoveLew,Rovebook,Thiebook}. This action, besides being invariant under local Lorentz transformations and spacetime diffeomorphisms, is invariant under an internal gauge symmetry analogous to (\ref{gaugetrn})~\cite{MGCD2017}. The goal of this section is to extend such an internal gauge symmetry to account for scalar, Yang-Mills, and fermion fields. 

Let us first recall some aspects about the gauge symmetries of the Holst action with a cosmological term. The action is given by $S[e,\omega]=\int_{\mathcal{M}^4} L_{\rm Holst}$, with the Lagrangian 4-form
\begin{eqnarray}\label{holst1}
L_{\rm Holst} = \kappa \left ( P_{IJKL} e^I\wedge e^J \wedge R^{KL}  - 2\Lambda \eta \right),
\end{eqnarray}
where $P_{IJKL}:= (1/2) \epsilon_{IJKL} + (\sigma/\gamma) \eta_{[I|K} \eta_{|J]L}$ and $\gamma\in\mathbb{R}-\{0\}$ is the Immirzi parameter~\cite{Barbero, Immirzi} (see~\cite{Montesinos2018} for the Hamiltonian formulation of the Holst action in terms of manifestly Lorentz-covariant variables and first-class constraints only). Note that (\ref{holst1}) differs from (\ref{palatinindim}) with $n=4$ by the Holst term $e^I\wedge e^J\wedge R_{IJ}$, which defines a topological field theory~\cite{Liu}. The variational derivatives coming from the action defined by (\ref{holst1}) are
\numparts
\begin{eqnarray}
{\mathcal E}_I := \frac{\delta S}{\delta e^I} = - 2 \kappa e^J \wedge \left( P_{IJKL} R^{KL} - \frac{\Lambda}{6} \epsilon_{IJKL}e^K\wedge e^L\right),\label{eqmoth1} \\
{\mathcal E}_{IJ} :=\frac{\delta S}{\delta \omega^{IJ}}= - \kappa D (P_{IJKL} e^K \wedge e^L). \label{eqmoth2}
\end{eqnarray}
\endnumparts
On-shell, ${\mathcal E}_I$ and ${\mathcal E}_{IJ}$ do not depend on the Immirzi parameter $\gamma$ and lead to Einstein's equations with a cosmological constant.

It was shown in~\cite{MGCD2017} that the infinitesimal internal transformation
\begin{eqnarray}
\delta_{\rho}e^I = && D \rho^I, \nonumber\\
\delta_{\rho} \omega^{IJ} = && Z^{IJ}{}_{KL}\rho^K e^L, \label{tresholstsym} 
\end{eqnarray}
where $Z^{IJ}{}_{\!KL}$ is given by
\begin{eqnarray}\label{Zholst}
\fl Z^{IJ}{}_{\!KL} \!:=\!	(P^{-1})^{IJPQ} \! \left [ \frac12 P_{KLMN} \mathcal{R}^{MN}{}_{\!\!PQ} \! -\! P_{KPMN} \mathcal{R}^{MN}{}_{\!\!QL} \! + \! \frac{\Lambda}{3}(P_{KLPQ} + \!2 P_{KPQL}) \right ]\!,\nonumber\\
\end{eqnarray}
is a gauge symmetry of the Holst action with a cosmological term. Here, we have defined $(P^{-1})^{IJKL}:=\sigma \gamma^2 (\gamma^2-\sigma)^{-1} [ (1/2) \epsilon^{IJKL} - (\sigma/\gamma) \eta^{[I|K} \eta^{|J]L}]$ such that $(P^{-1})^{IJKL}P_{KLMN}=\delta^I_{[M}\delta^J_{N]}$. It turns out that diffeomorphisms can be constructed out of local Lorentz transformations and the transformation~(\ref{tresholstsym}), modulo terms involving ${\mathcal E}_I$ and ${\mathcal E}_{IJ}$. Then, the full gauge invariance of general relativity can be equivalently described by the set of symmetries composed of either local Lorentz transformations and diffeomorphisms, or local Lorentz transformations and the transformation~(\ref{tresholstsym}).

Note that the transformation~(\ref{tresholstsym}) depends on the Immirzi parameter through $(P^{-1})^{IJKL}$ and $P_{IJKL}$, and that in the limit $\gamma\rightarrow\infty$ it reduces to~(\ref{gaugetrn}) with $n=4$ since  $Z^{IJ}{}_{KL}|_{\gamma\rightarrow\infty}=Z_4{}^{IJ}{}_{KL}$. Thus, in general,~(\ref{tresholstsym}) is different from~(\ref{gaugetrn}) with $n=4$, although the Holst Lagrangian and the four-dimensional Palatini Lagrangian lead to the same space of solutions of the equations of motion if the tetrad is non-degenerate. On the other hand, the algebra described by local Lorentz transformations and (\ref{tresholstsym}) is still open and given by  (\ref{algebrapalatnd}) but with $Z^{IJ}{}_{KL}$ instead of $Z_4{}^{IJ}{}_{KL}$  and the variational derivatives those given in (\ref{eqmoth1}) and (\ref{eqmoth2})~\cite{MGCD2017}. Furthermore, the off-shell identities~(\ref{GR-Id-LL}) and~(\ref{GR-Id-LN}) hold (with the corresponding variational derivatives), and the analog of~(\ref{GR-Id-NN}) reads
\begin{eqnarray}
\fl {\mathcal E}_I \wedge \left[ \delta_{\rho_1}, \delta_{\rho_2} \right] e^I + {\mathcal E}_{IJ} \wedge \left[ \delta_{\rho_1}, \delta_{\rho_2} \right] \omega^{IJ}  +  d\left[ -\tau^{IJ} {\mathcal E}_{IJ} \right. \nonumber \\ 
\fl \left. +  2 \kappa \rho^{[M}_1 \rho^{N]}_2 \left(  2P_{IJSK}  Z_M{}^{I}{}_N{}^{J} + 2P_{IJMK}  Z^{IJ}{}_{NS} + \Lambda \epsilon_{MNKS} \right)  e^S \wedge De^K \right]= 0, \label{Holst-Id}
\end{eqnarray}
with $\tau^{IJ}:=2 Z_K{}^{[I}{}_L{}^{J]} \rho^{[K}_1 \rho^{L]}_2 $. 

The following sections are devoted to studying the modifications induced on the internal gauge symmetry~(\ref{tresholstsym}) by the coupling of scalar, Yang-Mills, and fermion fields to the Holst action.

\subsection{Scalar field}

We consider here a real scalar field minimally coupled to the Holst action with a cosmological constant. The Lagrangian of the theory is
\begin{eqnarray}\label{Holst-sf-Lag}
L = L_{\rm Holst}  + \alpha\left[d\phi\wedge\star d\phi - V(\phi) \eta  \right],
\end{eqnarray}
where $V$ is an arbitrary function depending on the scalar field~$\phi$ and $\alpha$ is a real parameter. The variational derivatives of the action defined by this Lagrangian are
 \numparts
\begin{eqnarray}
 E_I :=\frac{\delta S}{\delta e^I}= {\mathcal E}_I -  T_{JI}\star e^J, \label{Holst-sf-Eqmot1} \\
 E_{IJ} :=\frac{\delta S}{\delta \omega^{IJ}}= {\mathcal E}_{IJ}, \label{Holst-sf-Eqmot2}\\
 E_{\phi} :=\frac{\delta S}{\delta \phi} = -2\alpha d(\star d\phi)-\alpha \frac{dV}{d\phi}\eta, \label{Holst-sf-Eqmot3}
\end{eqnarray}
\endnumparts
 where  ${\mathcal E}_I$ and ${\mathcal E}_{IJ}$ are given by (\ref{eqmoth1}) and (\ref{eqmoth2}), respectively, and
 \begin{eqnarray}\label{Holst-sf-EMtensor}
 T_{IJ}:=\alpha\left[-2\partial_I\phi\partial_J\phi+\left(\partial_K\phi\partial^K\phi-V\right)\eta_{IJ}\right],
 \end{eqnarray}
for $\partial_I\phi:=\partial_I\intprod d\phi$, is the symmetric energy-momentum tensor of the scalar field. It can be shown that, on-shell, (\ref{Holst-sf-Eqmot1}), (\ref{Holst-sf-Eqmot2}), and (\ref{Holst-sf-Eqmot3}) imply Einstein's equations with a cosmological constant and a minimally coupled scalar field.  

In the purpose of obtaining the internal gauge symmetry of the same kind of (\ref{tresholstsym}) under which the Lagrangian (\ref{Holst-sf-Lag}) is quasi-invariant, our starting point is to compute the covariant derivative of $ E_I$. In doing so, we obtain for the first and second terms on the r.h.s. of~(\ref{Holst-sf-Eqmot1})
\begin{eqnarray}
D\mathcal{E}_I= - E_{KL}\wedge Z{}^{KL}{}_{IJ}e^J, \label{Holst-sf-DEI1}
\end{eqnarray}
with $Z^{IJ}{}_{KL}$ given by (\ref{Zholst}), and
\begin{eqnarray}
D\left(T_{IJ}\star e^J\right) = E_{KL} \wedge S{}^{KL}{}_{IJ} e^J+E_{\phi}\partial_I\phi, \label{Holst-sf-DEI2}
\end{eqnarray}
where
\begin{eqnarray}\label{Holst-sf-S}
 S^{IJ}{}_{KL} &:=& -\frac{1}{4}\alpha\kappa^{-1}(P^{-1})^{IJPQ} \left[ \epsilon_{KLPQ} \left(\partial_N\phi\partial^N\phi-V\right) \right. \nonumber\\
 &&\left.- 4 \epsilon_{KLPN} \partial^N\phi\partial_Q\phi \!+ 2 \epsilon_{KPQN} \partial^N\phi\partial_L\phi  \right],
 \end{eqnarray}
respectively. Using (\ref{Holst-sf-DEI1}) and (\ref{Holst-sf-DEI2}), it is straightforward to see that
\begin{eqnarray}
DE_I=-E_{KL} \wedge \left(Z^{KL}{}_{IJ} + S{}^{KL}{}_{IJ} \right)e^J-E_{\phi}\partial_I\phi, \nonumber
\end{eqnarray}
or
\begin{eqnarray}\label{Holst-sf-IdSym}
DE_I+E_{KL} \wedge \left(Z^{KL}{}_{IJ} + S{}^{KL}{}_{IJ} \right)e^J+E_{\phi}\partial_I\phi=0,
\end{eqnarray}
which, after being multiplied by the parameter $\rho^I$ and rearranged, results in the off-shell identity 
\begin{eqnarray}\label{Holst-sf-IdSym2}
\fl E_I \wedge \underbrace{D\rho^I }_{\delta_{\rho} e^I}+ E_{IJ} \wedge \underbrace{\left(Z{}^{IJ}{}_{KL} + S{}^{IJ}{}_{KL}\right)\rho^K e^L}_{\delta_{\rho} \omega^{IJ}} + E_{\phi} \wedge \underbrace{\rho^I\partial_I \phi }_{\delta_{\rho}\phi} + d \left ( \rho^I E_I\right )=0.
\end{eqnarray}
  
It follows from this identity and the converse of Noether's second theorem that the infinitesimal internal transformation associated to the Noether identity~(\ref{Holst-sf-IdSym}), namely 
 \begin{eqnarray}\label{Holst-sf-newtr}
 & \delta_{\rho} e^I=D\rho^I,\nonumber\\
 & \delta_{\rho}\omega^{IJ}=\left(Z^{IJ}{}_{KL}+S{}^{IJ}{}_{KL}\right)\rho^K e^L,\nonumber\\
 & \delta_{\rho}\phi=\rho^I\partial_I \phi,
 \end{eqnarray}
is a gauge symmetry of the Holst action with a cosmological constant and a minimally coupled scalar field. It can be checked that~(\ref{Holst-sf-newtr}) leaves the Lagrangian (\ref{Holst-sf-Lag}) quasi-invariant because it changes as
 \begin{eqnarray}
\fl 	\delta_{\rho} L= d \left[ \kappa \rho^I e^J \wedge \left(P_{IJKL} R^{KL} + \frac{\Lambda}{6} \epsilon_{IJKL}e^K\wedge e^L\right) +\frac{1}{2} \rho^I (T_{IJ}+2\alpha V \eta_{IJ}) \star e^J \right].
\end{eqnarray}

Notice that the transformation of the frame in (\ref{Holst-sf-newtr}), which is the same as in~(\ref{tresholstsym}), is not affected by the presence of the scalar field. In contrast, the transformation of the connection contains the extra term $S{}^{IJ}{}_{KL}$, which depends on this matter field and can be rewritten as  
 \begin{eqnarray}
S^{IJ}{}_{KL} =-\frac{1}{4}\kappa^{-1}(P^{-1})^{IJPQ} \left[2 \epsilon_{KLPN} T^N{}_Q - \epsilon_{KPQN} T^N{}_L  \right], \label{Holst-sf-S2}
\end{eqnarray}
where $T_{IJ}$ is given by (\ref{Holst-sf-EMtensor}). Finally, we find that the analog of the off-shell identity~(\ref{Holst-Id}) reads
\begin{eqnarray}
\fl {\mathcal E}_I \wedge \left [ \delta_{\rho_1}, \delta_{\rho_2} \right ] e^I + {\mathcal E}_{IJ} \wedge \left [ \delta_{\rho_1}, \delta_{\rho_2} \right ] \omega^{IJ} + {\mathcal E}_{\phi}  \left [ \delta_{\rho_1}, \delta_{\rho_2} \right ] \phi +  d\Big\{ - \tau^{IJ} {\mathcal E}_{IJ}  \nonumber  \\
\fl + 2 \kappa \rho^{[M}_1 \rho^{N]}_2  \left[  P_{IJSK} \left(2 Z_M{}^{I}{}_N{}^{J} + 2 S_M{}^{I}{}_N{}^{J} - S^{IJ}{}_{MN} \right)    \right.   \nonumber \\
\fl \left.  \Big. +  P_{IJMK} \left( 2 Z^{IJ}{}_{NS} + S^{IJ}{}_{NS} \right) +  \Lambda \epsilon_{MNKS}    \Big] e^S \wedge De^K \right\}=0, \label{Holst-sf-Id}
\end{eqnarray}
where $\tau^{IL}:=2 \left(Z_J{}^{[I}{}_K{}^{L]}+S{}_J{}^{[I}{}_K{}^{L]}\right)  \rho^{[J}_1 \rho^{K]}_2$.



 \subsection{Yang-Mills field}

We will now focus on the minimal coupling of a Yang-Mills field to the Holst action with a cosmological constant. Therefore, the Lagrangian under consideration here is
 \begin{eqnarray}\label{Holst-YM-Lag}
  L = L_{\rm Holst} + \alpha \star F_a \wedge F^a.
 \end{eqnarray}
The notation related to the Yang-Mills field is the same as in section~\ref{sec-nPal-D}. From the action defined by this Lagrangian, we obtain the following variational derivatives
\numparts
\begin{eqnarray}
 E_I :=\frac{\delta S}{\delta e^I} =  {\mathcal E}_I -  T_{JI}\star e^J, \label{Holst-YM-Eqmot1} \\
E_{IJ} :=\frac{\delta S}{\delta \omega^{IJ}}= {\mathcal E}_{IJ}, \label{Holst-YM-Eqmot2}\\
E_{a} := \frac{\delta S}{\delta A^a} = -2 \alpha \mathcal{D}\star F_a , \label{Holst-YM-Eqmot3}
 \end{eqnarray}
 \endnumparts
where  ${\mathcal E}_I$ and ${\mathcal E}_{IJ}$ are, respectively, given by (\ref{eqmoth1}) and (\ref{eqmoth2}), whereas
 \begin{eqnarray}\label{Holst-YM-EMtensor}
 T_{IJ}:=-2\alpha\left( F_{aIK} F^{a}{}_{J}{}^{K} -\frac{1}{4} F_{aKL} F^{aKL}  \eta_{IJ}\right),
 \end{eqnarray}
is the  symmetric energy-momentum tensor. On-shell, $E_I$, $E_{IJ}$, and $E_{a}$ are equivalent to the equations of motion coming from the action defined by (\ref{nPal-YM-Lag}) with $n=4$. Hence, the Immirzi parameter $\gamma$ does not change the classical dynamics.

Applying the procedure, we start by computing $DE_I$. The covariant derivative of ${\mathcal E}_I$ in (\ref{Holst-YM-Eqmot1}) leads to (\ref{Holst-sf-DEI1}) with $E_{IJ}$ given by~(\ref{Holst-YM-Eqmot2}), whereas for the second term of (\ref{Holst-YM-Eqmot1}) we have
\begin{eqnarray}\label{Holst-YM-DEI2}
D\left(T_{JI}\star e^J\right)=E_{KL} \wedge Q^{KL}{}_{IJ}e^J + E_{a} \wedge F^a{}_{IJ} e^J,
\end{eqnarray}
with
\begin{eqnarray}
\fl  Q^{IJ}{}_{KL} &:=&-\frac{1}{4}\alpha\kappa^{-1} (P^{-1})^{IJPQ} \left[ \frac12\epsilon_{PQKL} F_{aMN} F^{aMN} -2 \epsilon_{LPMN} F_{aKQ} F^{aMN}\right. \nonumber\\
\fl &&\Big. + \epsilon_{PQMN} F_{aKL} F^{aMN} - 2 \epsilon_{KLMN} F_{aPQ} F^{aMN} \Big].\label{Holst-YM-Q}
\end{eqnarray}
In the process of obtaining (\ref{Holst-YM-DEI2}), we used the Bianchi identity $\mathcal{D}F^a=0$. Combining these results, we end up with
\begin{eqnarray} \label{Holst-YM-IdSym1}
DE_I=-E_{KL} \wedge (Z^{KL}{}_{IJ}+Q^{KL}{}_{IJ}) e^J-E_{a} \wedge F^a{}_{IJ} e^J, \nonumber
\end{eqnarray}
or
\begin{eqnarray} \label{Holst-YM-IdSym}
DE_I+E_{KL} \wedge (Z^{KL}{}_{IJ}+Q^{KL}{}_{IJ}) e^J+E_{a} \wedge F^a{}_{IJ} e^J =0.
\end{eqnarray}
If we now multiply (\ref{Holst-YM-IdSym}) by the parameter $\rho^I$ and rearrange it, then we arrive at the off-shell identity 
\begin{eqnarray}\label{Holst-YM-IdSym2}
\fl E_I \wedge \underbrace{D\rho^I }_{\delta_{\rho} e^I}+ E_{IJ} \wedge \underbrace{\left(Z^{IJ}{}_{KL} + Q^{IJ}{}_{KL}\right)\rho^K e^L}_{\delta_{\rho} \omega^{IJ}} + E_{a} \wedge \underbrace{F^a{}_{IJ} \rho^I e^J }_{\delta_{\rho}A^a} + d \left(\rho^I {\mathcal E}_I\right)=0.
\end{eqnarray} 

By virtue of the converse of Noether's second theorem, from (\ref{Holst-YM-IdSym2}) we have that the infinitesimal internal transformation
\begin{eqnarray}\label{Holst-YM-newtr}
& \delta_{\rho} e^I=D\rho^I,\nonumber\\
& \delta_{\rho}\omega^{IJ}=\left(Z^{IJ}{}_{KL} + Q^{IJ}{}_{KL} \right) \rho^K e^L,\nonumber\\
& \delta_{\rho}A^a=F^a{}_{IJ}\rho^I e^J,
\end{eqnarray}
is a gauge symmetry of the Holst action with a cosmological constant and a minimally coupled Yang-Mills field. In fact, it can be readily checked that the Lagrangian~(\ref{Holst-YM-Lag}) is quasi-invariant under~(\ref{Holst-YM-newtr}), since
\begin{eqnarray}
\delta_{\rho} L &=& d \left[ \kappa \rho^I e^J \wedge \left(P_{IJKL} R^{KL} + \frac{\Lambda}{6} \epsilon_{IJKL}e^K\wedge e^L\right) \right.\nonumber \\ 
&&\left.+ \frac{1}{2} \rho^I (T_{IJ}+\alpha F_{aKL} F^{aKL} \eta_{IJ}) \star e^J \right].
\end{eqnarray}
In addition, it is worth pointing out that (\ref{Holst-YM-IdSym}) is the Noether identity associated to the gauge transformation~(\ref{Holst-YM-newtr}).

From (\ref{Holst-YM-newtr}) it is clear that the transformations of the frame and the Yang-Mills connection are the same as in (\ref{nPal-YM-newtr}), while the transformation of the connection is modified by both the Immirzi parameter and the Yang-Mills field. It is interesting to note that $Q^{IJ}{}_{KL}$ can be written in terms of the symmetric energy-momentum tensor~(\ref{Holst-YM-EMtensor}) as
\begin{eqnarray}
Q^{IJ}{}_{KL} =-\frac{1}{4}\kappa^{-1}(P^{-1})^{IJPQ} \left[2 \epsilon_{KLPN} T^N{}_Q - \epsilon_{KPQN} T^N{}_L  \right], \label{Holst-YM-Q2}
\end{eqnarray}
which resembles its analog~(\ref{Holst-sf-S2}) for the case of the scalar field. Also, as it has to be, when $\gamma\rightarrow\infty$ the gauge transformation (\ref{Holst-YM-newtr}) collapses to (\ref{nPal-YM-newtr}) with $n=4$. 
 
 \subsection{Fermions}\label{subsec:merc}

In this section, we consider the coupling of fermions to gravity  described by the action proposed in~\cite{Mercuri2006}, whose Lagrangian is given by
\begin{eqnarray}\label{Mer-D-Lag}
\fl L = L_{\rm Holst}   + \frac{1}{2} \left[\overline{\psi}  \gamma^I \left( \mathbb{I} - \frac{\rmi}{\alpha} \gamma^5\right) D\psi - \overline{D\psi} \left( \mathbb{I} - \frac{\rmi}{\alpha}\gamma^5 \right) \gamma^I \psi \right]\wedge \star e_I -  m\overline{\psi} \psi \eta,
\end{eqnarray}
where $\gamma^5:=\rmi \gamma^0 \gamma^1 \gamma^2 \gamma^3$ and $\alpha$ is a coupling parameter. Here, we restrict ourselves to Lorentzian ($\sigma=-1$) manifolds $\mathcal{M}^4$ and use the same notation introduced in section~\ref{sec-nPal-D} for the fermionic matter. This Lagrangian is used to study parity violation effects in quantum gravity~\cite{BojowaldDas2008}. When $\alpha\rightarrow\infty$, it describes general relativity with minimally coupled fermions in the presence of the Immirzi parameter, which in this case has physical implications~\cite{Freidel2005,Perez-Rovelli-2006}. However, if instead of that limit we take $\alpha=-\gamma$, it turns out that  the effective action of the theory obtained by integrating out the connection is equal (modulo a total derivative involving the Immirzi parameter) to the one obtained from the Palatini action with minimally coupled fermions~\cite{Mercuri2006}. For $\alpha \neq -\gamma$, the Lagrangian (\ref{Mer-D-Lag}) describes fermions non-minimally coupled to gravity.

The variational derivatives that emerge from the action defined by (\ref{Mer-D-Lag}) are
\numparts
\begin{eqnarray}
\fl E_I &:=& \frac{\delta S}{\delta e^I} = {\mathcal E}_I - T_{JI}\star e^J , \label{Mer-D-Eqmot1} \\
\fl E_{IJ} &:=&  \frac{\delta S}{\delta \omega^{IJ}} =   {\mathcal E}_{IJ} + \frac{1}{4} \epsilon_{IJKL} J^K \star e^L + \frac{1}{2\alpha}  J_{[I} \star e_{J]}, \label{Mer-D-Eqmot2}\\
\fl E_{\psi} &:=& \frac{\delta S}{\delta \psi} = - \overline{D\psi}  \gamma^I \wedge \star e_I - \frac{1}{2} \overline{\psi} \gamma^I D\star e_I -  m \overline{\psi} \eta + \frac{\rmi}{2\alpha} \overline{\psi} \gamma_I \gamma^5 D\star e^I, \label{Mer-D-Eqmot3} \\
\fl E_{\overline{\psi}} &:=& \frac{\delta S}{\delta \overline{\psi}} =  \gamma^I D\psi   \wedge \star e_I + \frac{1}{2} \gamma^I \psi  D\star e_I -  m \psi \eta- \frac{\rmi}{2 \alpha} \gamma^5 \gamma_I \psi D \star e^I , \label{Mer-D-Eqmot4}
\end{eqnarray}
\endnumparts
where ${\mathcal E}_I$ and ${\mathcal E}_{IJ}$ are, respectively, the same as those given in (\ref{eqmoth1}) and (\ref{eqmoth2});
\begin{eqnarray}\label{Mer-D-EMtensor}
T_{IJ}&:=&-\frac{1}{2} \left( \overline{\psi} \gamma_I D_J \psi - \overline{D_J \psi} \gamma_I \psi + \frac{1}{\alpha} D_J J_I\right) \nonumber \\
&&+ \frac12 \left(  \overline{\psi} \gamma^K D_K \psi - \overline{D_K \psi} \gamma^K \psi +\frac{1}{\alpha} D_K J^K - 2m \overline{\psi} \psi \right) \eta_{IJ},
\end{eqnarray}
is the nonsymmetric energy-momentum tensor; and $J^I=\rmi \overline{\psi} \gamma^5 \gamma^I \psi$ is the axial fermion current. Furthermore, in (\ref{Mer-D-Eqmot2}) we have used $\{\gamma_I, \sigma_{JK}\}=\rmi \epsilon_{IJKL} \gamma^5 \gamma^L$. Here we see that,  in the limit $\alpha\rightarrow\infty$, the equations of motion $E_I=0$  and $E_{IJ}=0$ depend on the  Immirzi parameter, which then affects the classical dynamics. Nevertheless, if we take $\alpha=-\gamma$, the resulting equations of motion obtained by setting (\ref{Mer-D-Eqmot1})-(\ref{Mer-D-Eqmot4}) equal to zero are the {\it same} than those obtained from the Palatini action with minimally coupled Dirac fermions and obtained by setting (\ref{nPal-D-Eqmot1})-(\ref{nPal-D-Eqmot4}) equal to zero, i.e., the Immirzi parameter $\gamma$ drops out from the equations of motion coming from (\ref{Mer-D-Eqmot1})-(\ref{Mer-D-Eqmot4}). That is to say, in the same sense that the Immirzi parameter drops out from the equations of motion coming from the Holst action leading to the equations of motion obtained from the Palatini action, for $\alpha=-\gamma$ the Immirzi parameter drops out when (\ref{Mer-D-Eqmot1})-(\ref{Mer-D-Eqmot4}) are equal to zero leading to (\ref{nPal-D-Eqmot1})-(\ref{nPal-D-Eqmot4}) equal to zero.

We now turn to exhibit the analog of the internal gauge symmetry~(\ref{tresholstsym}) for the action defined by~(\ref{Mer-D-Lag}). Following the recipe, we need to compute the covariant derivative of Eq.~(\ref{Mer-D-Eqmot1}). The result for the first term on the r.h.s. of this equation is
\begin{eqnarray}\label{Mer-D-DEI1}
\fl	D\mathcal{E}_I= - E_{KL}\wedge Z{}^{KL}{}_{IJ}e^J -\frac{1}{4} \left( \epsilon_{KL}{}^{NJ} + \frac{2}{\alpha} \delta^{[N}_{K} \delta^{J]}_{L}   \right) J_{N} Z{}^{KL}{}_{IJ}\, \eta,
\end{eqnarray}
while for the second one is
\begin{eqnarray}\label{Mer-D-DEI2}
\fl	D\left(T_{JI}\star e^J\right) &=& E_J \wedge \left\{ \frac{\gamma}{4\kappa\alpha(\gamma^2+1)}\left[ (1-\alpha \gamma) \epsilon^{JK}{}_{IL} + (\alpha+\gamma)  \delta^{[J}_{I} \delta^{K]}_{L}  \right] J_K e^L\right\} \nonumber \\
\fl &&+ E_{KL}\wedge F^{KL}{}_{IJ}e^J + E_{\psi} D_I \psi + \overline{D_I \psi} E_{\overline{\psi}} \nonumber \\
\fl && -\frac{1}{4} \left( \epsilon_{KL}{}^{NJ} + \frac{2}{\alpha} \delta^{[N}_{K} \delta^{J]}_{L}   \right) J_{N} Z{}^{KL}{}_{IJ}\,\eta,
\end{eqnarray}
with
\begin{eqnarray}\label{Mer-D-Q}
\fl	F^{IJ}{}_{KL} &:=& -\frac{1}{4}\kappa^{-1} (P^{-1})^{IJPQ} \left[ \frac{1}{2} \epsilon_{KLPQ} \left(  \overline{\psi} \gamma^N D_N \psi - \overline{D_N \psi} \gamma^N \psi +\frac{1}{\alpha} D_N J^N - 2m \overline{\psi} \psi \right) \right. \nonumber \\
\fl &&-\epsilon_{KLPN} \left( \overline{\psi} \gamma^N D_Q \psi - \overline{D_Q \psi} \gamma^N \psi + \frac{1}{\alpha} D_Q J^N\right) \nonumber \\
\fl &&\left.+ \frac{1}{2} \epsilon_{KPQN} \left( \overline{\psi} \gamma^N D_L \psi - \overline{D_L \psi} \gamma^N \psi + \frac{1}{\alpha} D_L J^N\right) \right].
\end{eqnarray}
Once these results are put together, we obtain
\begin{eqnarray}
	DE_I = - E_J \wedge \left\{ \frac{\gamma}{4\kappa\alpha(\gamma^2+1)}\left[ (1-\alpha \gamma) \epsilon^{JK}{}_{IL} + (\alpha+\gamma)  \delta^{[J}_{I} \delta^{K]}_{L}  \right] J_K e^L\right\}  \nonumber \\
	- E_{KL} \wedge \left(Z^{KL}{}_{IJ} + F^{KL}{}_{IJ} \right)e^J - E_{\psi} D_I \psi - \overline{D_I \psi} E_{\overline{\psi}}, \nonumber
\end{eqnarray}
or
\begin{eqnarray}\label{Mer-D-IdSym}
	DE_I + E_J \wedge \left\{ \frac{\gamma}{4\kappa\alpha(\gamma^2+1)}\left[ (1-\alpha \gamma) \epsilon^{JK}{}_{IL} + (\alpha+\gamma)  \delta^{[J}_{I} \delta^{K]}_{L}  \right] J_K e^L\right\}  \nonumber \\
	+ E_{KL} \wedge \left(Z^{KL}{}_{IJ} + F^{KL}{}_{IJ} \right)e^J + E_{\psi} D_I \psi + \overline{D_I \psi} E_{\overline{\psi}}   =0.
\end{eqnarray}
After multiplying (\ref{Mer-D-IdSym}) by the gauge parameter $\rho^I$ and rearranging, the resulting off-shell identity is 
\begin{eqnarray}\label{Mer-D-IdSym2}
\fl E_I \wedge \underbrace{\left\{D\rho^I+ \frac{\gamma}{4\kappa\alpha(\gamma^2+1)} \left[ (1-\alpha \gamma) \epsilon^{I}{}_{JKL}  J^J \rho^{K} e^{L} + (\alpha+\gamma) J_J \rho^{[I} e^{J]} \right]  \right\} }_{\delta_{\rho} e^I} \nonumber \\
\fl + E_{IJ} \wedge \underbrace{\left(Z^{IJ}{}_{KL} + F^{IJ}{}_{KL}\right)\rho^K  e^L}_{\delta_{\rho} \omega^{IJ}} + E_{\psi} \underbrace{\rho^I D_I \psi}_{\delta_{\rho} \psi}  + \underbrace{ \rho^I \overline{D_I \psi} }_{\delta_{\rho} \overline{\psi}} E_{\overline{\psi}} + d \left( \rho^I E_I\right)=0.
\end{eqnarray}

By appealing to the converse of Noether's second theorem again, (\ref{Mer-D-IdSym2}) allows us to read off the infinitesimal internal transformation
\begin{eqnarray}\label{Mer-D-newtr}
\fl & \delta_{\rho} e^I=D\rho^I+ \frac{\gamma}{4\kappa\alpha(\gamma^2+1)} \left[ (1-\alpha \gamma) \epsilon^{I}{}_{JKL}  J^J \rho^{K} e^{L} + (\alpha+\gamma) J_J \rho^{[I} e^{J]} \right],\nonumber\\
\fl & \delta_{\rho}\omega^{IJ}=\left(Z^{IJ}{}_{KL}+F^{IJ}{}_{KL}\right)\rho^K e^L,\nonumber\\
\fl & \delta_{\rho} \psi=\rho^I D_I \psi,  \nonumber\\
\fl & \delta_{\rho} \overline{\psi}=\rho^I \overline{D_I \psi}, 
 \end{eqnarray} 
which is a gauge symmetry of the action defined by~(\ref{Mer-D-Lag}) and is associated to the Noether identity~(\ref{Mer-D-IdSym}). Certainly, it can be directly checked that under (\ref{Mer-D-newtr}) the Lagrangian (\ref{Mer-D-Lag}) is quasi-invariant because 
 \begin{eqnarray}\label{Mer-D-changeL}
\delta_{\rho} L &=& d\left\{\kappa \rho^I e^J \wedge \left(P_{IJKL} R^{KL} + \frac{\Lambda}{6} \epsilon_{IJKL}e^K\wedge e^L\right) \right.\nonumber \\ 
&&\left.+ \frac{1}{2} \rho^I \left[T_{JI} -\frac{1}{3} \left( T^K{}_K -2m\overline{\psi} \psi \right) \eta_{IJ}\right] \star e^J \right\}.
\end{eqnarray}

As in the case of (\ref{nPal-D-newtr}), the coupling of fermionic matter to gravity generates modifications in the transformation of the frame~$\delta_{\rho} e^I$ in~(\ref{Mer-D-newtr}). Such modifications are characterized by the axial fermion current $J^I$, and involve both the Immirzi parameter $\gamma$ and the parameter $\alpha$. Moreover, as expected, the transformation of the connection $\delta_{\rho}\omega^{IJ}$ is also affected by the matter field; it involves the extra term $F^{IJ}{}_{KL}$ coming from the presence of the fermion field. It is worth noting here that once we write down $F^{IJ}{}_{KL}$ in terms of the nonsymmetric energy-momentum tensor (\ref{Mer-D-EMtensor}), namely
\begin{eqnarray}\label{Mer-D-Q2}
F^{IJ}{}_{KL} = -\frac{1}{4}\kappa^{-1} (P^{-1})^{IJPQ} \left[2 \epsilon_{KLPN} T^N{}_Q - \epsilon_{KPQN} T^N{}_L  \right],
\end{eqnarray}
it takes the same structural form as that of (\ref{Holst-sf-S2}) or (\ref{Holst-YM-Q2}). Regarding the transformations $\delta_{\rho} \psi$ and $\delta_{\rho} \overline{\psi}$, we can see that they remain unchanged with respect to those of~(\ref{nPal-D-newtr}). Finally, let us consider two particular cases of the gauge symmetry~(\ref{Mer-D-newtr}). On one hand, when $\gamma\rightarrow\infty$ and $\alpha\rightarrow\infty$, we have that (\ref{Mer-D-newtr}) reduces to (\ref{nPal-D-newtr}) with $n=4$, which is expected since in this case the Lagrangian (\ref{Mer-D-Lag}) is nothing but the Palatini Lagrangian with minimally coupled fermions. On the other hand, if we set $\alpha=-\gamma$ in (\ref{Mer-D-newtr}), the Immirzi parameter disappears from the transformation of the frame $\delta_{\rho} e^I$, which becomes the first line of (\ref{nPal-D-newtr}) with $n=4$, but is still present through $F^{IJ}{}_{KL}$ in the transformation of the connection $\delta_{\rho}\omega^{IJ}$ given in~(\ref{Mer-D-newtr}).

\section{Conclusion}

In this paper, we have extended the higher-dimensional generalization of three-dimensional local translations, namely the internal gauge symmetry~(\ref{gaugetrn}), to address the minimal coupling of Yang-Mills and fermion fields to the $n$-dimensional Palatini action with a cosmological constant. We have explicitly obtained the extended internal gauge symmetries for each one of the matter-gravity couplings from the direct application of the converse of Noether's second theorem. A notable feature of the resulting extended symmetry is that its structure is sensitive to both the spacetime dimension and the presence of the matter field. In particular, we have shown that the transformation of the connection encompassing the Yang-Mills field and the one encompassing the fermion field take the same form when written in terms of the respective energy-momentum tensors. This happens even though the energy-momentum tensors of the Yang-Mills and fermion fields are, respectively, symmetric and nonsymmetric. Also, in the case of fermions the transformation of the frame has an extra term, as compared to its analog in the case of Yang-Mills field, that in four dimensions is proportional to the axial fermion current. We did not discuss the extended symmetry for the coupling of a scalar field to the $n$-dimensional Palatini action since it has already been reported in~\cite{MGCD2017}. In this paper, we have also extended the analog of this internal gauge symmetry for the Holst action with a cosmological constant to include minimally coupled scalar and Yang-Mills fields, as well as the $\alpha$-parameter family of coupled fermions proposed in~\cite{Mercuri2006}. We found that, just as in the case of the $n$-dimensional Palatini action, the extended symmetry for the Holst action depends on the components of energy-momentum tensor, and for fermion fields, it has an additional modification characterized by the axial fermion current. Moreover, the internal gauge symmetry extended to include the coupling of fermions described in~\cite{Mercuri2006} involves the Immirzi parameter $\gamma$ and the parameter $\alpha$, and even when $\alpha=-\gamma$, in which case $\gamma$ drops out from the equations of motion, the (off-shell) extended symmetry still depends on the Immirzi parameter. It is worth mentioning that infinitesimal diffeomorphims can be expressed, irrespective of the action principle and matter fields considered, as linear combinations of the resulting extended gauge symmetry and local Lorentz transformations with field-dependent gauge parameters, up to terms proportional to the variational derivatives of the action under consideration. Thus, the extended gauge symmetry together with local Lorentz transformations comprise a fundamental set of internal symmetries to describe the full gauge symmetry of general relativity coupled to matter fields.

These results provide new insights into the  nature of the internal gauge symmetry of the Palatini action uncovered in~\cite{MGCD2017}. For the sake of completeness it would be desirable to obtain the gauge algebra involving the extended internal symmetry (for each matter field) and local Lorentz transformations. In this regard, since the commutator algebra among (\ref{gaugetrn}) and local Lorentz transformations is open, we expect the same to be true for the resulting gauge algebra of the symmetries including the matter fields. Also, considering the relevance of the Immirzi parameter at the quantum level, it would be of interest to study the extension of the results of section~{\ref{subsec:merc}} to the general non-minimal coupling of fermions introduced in~\cite{Alexandrov2008}.  

\ack 

This work was partially supported by Consejo Nacional de Ciencia y Tecnolog\'{i}a (CONACyT), M\'{e}xico, Grants Nos. 237004-F and 237351. Mariano Celada would like to thank the financial support of Programa para el Desarrollo Profesional Docente, para el Tipo Superior (PRODEP) Grant No. 12313153 (through UAM-I). Diego Gonzalez is supported with a DGAPA-UNAM postdoctoral fellowship.

\section*{References}
\bibliography{references}

\providecommand{\newblock}{}
\begin{thebibliography}{10}
\expandafter\ifx\csname url\endcsname\relax
  \def\url#1{{\tt #1}}\fi
\expandafter\ifx\csname urlprefix\endcsname\relax\def\urlprefix{URL }\fi
\providecommand{\eprint}[2][]{\url{#2}}

\bibitem{Sotiriou2010}
Sotiriou T~P and Faraoni V 2010 {\em Rev. Mod. Phys.\/} {\bf
  \href{https://link.aps.org/doi/10.1103/RevModPhys.82.451}{82}} 451

\bibitem{Utiyama}
Utiyama R 1956 {\em Phys. Rev.\/} {\bf
  \href{http://link.aps.org/doi/10.1103/PhysRev.101.1597}{101}} 1597

\bibitem{Thiebook}
Thiemann T 2007 {\em Modern Canonical Quantum General Relativity\/} (Cambridge
  University Press, Cambridge)

\bibitem{AchTow}
Ach\'ucarro A and Townsend P 1986 {\em Phys. Lett. B\/} {\bf
  \href{http://www.sciencedirect.com/science/article/pii/0370269386901401}{180}}
  89

\bibitem{Witten}
Witten E 1988 {\em Nucl. Phys. B\/} {\bf
  \href{http://dx.doi.org/10.1016/0550-3213(88)90143-5}{311}} 46

\bibitem{Carlip2+1}
Carlip S 2003 {\em Quantum Gravity in 2+1 Dimensions\/} (Cambridge University
  Press, Cambridge)

\bibitem{MGCD2017}
Montesinos M, Gonz\'alez D, Celada M and D\'iaz B 2017 {\em Class. Quantum
  Grav.\/} {\bf \href{http://stacks.iop.org/0264-9381/34/i=20/a=205002}{34}}
  205002

\bibitem{Noether}
Noether E 1918 {\em Nachr. d. K\"onig. Gesellsch. d. Wiss. zu G\"ottingen,
  Math-phys. Klasse\/} {\bf \href{http://eudml.org/doc/59024}{1918}} 235

\bibitem{Bessel-Hagen}
Bessel-Hagen E 1921 {\em Math. Ann.\/} {\bf
  \href{http://eudml.org/doc/158894}{84}} 258

\bibitem{EmmyNoether}
Noether E 1971 {\em Transp. Theory and Stat. Phys.\/} {\bf
  \href{http://dx.doi.org/10.1080/00411457108231446}{1}} 186

\bibitem{Holst}
Holst S 1996 {\em Phys. Rev. D\/} {\bf
  \href{http://link.aps.org/doi/10.1103/PhysRevD.53.5966}{53}} 5966

\bibitem{Mercuri2006}
Mercuri S 2006 {\em Phys. Rev. D\/} {\bf
  \href{https://link.aps.org/doi/10.1103/PhysRevD.73.084016}{73}} 084016

\bibitem{Gerardo}
Torres~del Castillo G~F 2012 {\em Differentiable Manifolds: A Theoretical
  Physics Approach\/} (Birkh{\"a}user, New York)

\bibitem{Henneaux}
Henneaux M and Teitelboim C 1992 {\em Quantization of Gauge Systems\/}
  (Princeton University Press, Princeton)

\bibitem{Henneaux199047}
Henneaux M 1990 {\em Nucl. Phys. B (Proc. Suppl.)\/} {\bf
  \href{http://www.sciencedirect.com/science/article/pii/092056329090647D}{18}}
  47

\bibitem{RoveLew}
Ashtekar A and Lewandowski J 2004 {\em Class. Quantum Grav.\/} {\bf
  \href{http://stacks.iop.org/0264-9381/21/i=15/a=R01}{21}} R53

\bibitem{Rovebook}
Rovelli C 2004 {\em Quantum Gravity\/} (Cambridge University Press, Cambridge)

\bibitem{Barbero}
Barbero~G J~F 1995 {\em Phys. Rev. D\/} {\bf
  \href{http://link.aps.org/doi/10.1103/PhysRevD.51.5507}{51}} 5507

\bibitem{Immirzi}
Immirzi G 1997 {\em Class. Quantum Grav.\/} {\bf
  \href{http://stacks.iop.org/0264-9381/14/i=10/a=002}{14}} L177

\bibitem{Montesinos2018}
Montesinos M, Romero J and Celada M 2018 {\em Phys. Rev. D\/} {\bf
  \href{https://link.aps.org/doi/10.1103/PhysRevD.97.024014}{97}} 024014

\bibitem{Liu}
Liu L, Montesinos M and Perez A 2010 {\em Phys. Rev. D\/} {\bf
  \href{http://link.aps.org/doi/10.1103/PhysRevD.81.064033}{81}} 064033

\bibitem{BojowaldDas2008}
Bojowald M and Das R 2008 {\em Phys. Rev. D\/} {\bf
  \href{https://link.aps.org/doi/10.1103/PhysRevD.78.064009}{78}} 064009

\bibitem{Freidel2005}
Freidel L, Minic D and Takeuchi T 2005 {\em Phys. Rev. D\/} {\bf
  \href{https://link.aps.org/doi/10.1103/PhysRevD.72.104002}{72}} 104002

\bibitem{Perez-Rovelli-2006}
Perez A and Rovelli C 2006 {\em Phys. Rev. D\/} {\bf
  \href{https://link.aps.org/doi/10.1103/PhysRevD.73.044013}{73}} 044013

\bibitem{Alexandrov2008}
Alexandrov S 2008 {\em Class. Quantum Grav.\/} {\bf
  \href{http://stacks.iop.org/0264-9381/25/i=14/a=145012}{25}} 145012

\end{thebibliography}

\end{document}